\documentclass[11pt]{article}

\PassOptionsToPackage{numbers, compress}{natbib}

\usepackage[preprint]{neurips_2019}

\usepackage[utf8]{inputenc} 
\usepackage[T1]{fontenc}    
\usepackage{hyperref}       
\usepackage{url}            
\usepackage{booktabs}       
\usepackage{amsfonts}       
\usepackage{nicefrac}       
\usepackage{microtype}      
\usepackage{graphicx}
\usepackage{float}
\usepackage[belowskip=-10pt,aboveskip=0pt]{caption}
\title{Pattern Ensembling for Spatial Trajectory Reconstruction}

\author{%
  Shivam Pathak$^1$\\
  \texttt{skp454@nyu.edu} \\
   \And
  Mingyi He$^1$\\
  \texttt{mh5172@nyu.edu} \\
   \And
  Sergey Malinchik$^2$\\
  \texttt{sergey.b.malinchik@lmco.com} \\
   \And
  Stanislav Sobolevsky$^1$\\
  \texttt{ss9872@nyu.edu} \\
  \\
  $^1$Center for Urban Science and Progress\\
  New York University\\
  New York, NY, 11201 \\
  $^2$Lockheed Martin Advanced Technology Lab\\
  Lockheed Martin Corporation\\
  Cherry Hill, NJ, 08002\\
}

\begin{document}

\maketitle

\begin{abstract}


Digital sensing provides unprecedented opportunity to assess and understand mobility. However incompleteness, missing information, possible inaccuracies and temporal heterogeneity in the geolocation data can undermine its applicability. As mobility patterns are often repeated, we propose a method to use similar trajectory patterns from local vicinity and probabilistically ensemble them to robustly reconstruct missing or unreliable observations. We evaluate the proposed approach in comparison with traditional functional trajectory interpolation using a case of sea vessel trajectory data provided by The Automatic Identification System (AIS). By effectively leveraging the similarities in real-world trajectories, our pattern ensembling method helps reconstructing missing trajectory segments of extended length and complex geometry. It can be used for locating mobile objects when temporary unobserved as well as for creating an evenly sampled trajectory interpolation useful for further trajectory mining.


\end{abstract}
\section{Introduction}



The recent advances in location acquisition technologies and the widespread use of digital devices has helped curating a variety of spatial trajectories data, representing the mobility patterns of various  objects, such as people, vehicles or vessels.
A large body of papers demonstrates inferences and modeling of human mobility based on partial information provided by opportunistic digital sensing, such as anonymized mobile phone data \cite{grauwin2017identifying,kung2014exploring,hoteit2014estimating}, GPS traces \cite{santi2014quantifying,nyhan2016predicting,khulbe2020transportation}, Bluetooth \cite{yoshimura2017analysis}, credit card transactions \cite{hashemian2017socioeconomic} as well as social media data \cite{hawelka2014geo,sobolevsky2018twitter} along with successful urban applications of such inferences. 

However incompleteness, missing information and possible inaccuracies in such data often undermine its applicability. Furthermore, trajectory mining can benefit from having regular representations of the trajectory, including having location observations at even time intervals, which is often not the case for the real-world trajectory data. In order to address this problem the present paper proposes an approach for inferring missing locations and the entire trajectory segments based on available partial observations. 

We will illustrate the approach on one of the examples of mobility data - The Automatic Identification System (AIS), being a rich source of spatial data for sea vessels which also serves as a representative illustration of the above issues.
The AIS is an automatic vessel tracking system used by Vessel Traffic Services(VTS) for maritime navigation and safety. The AIS uses transponders on the ship to transmit time-stamped vessel positional information to nearby ships and VTS to identify and locate vessels. It also collects dynamic data related to ships movements such as speed, time, and positional coordinates. The dynamic data of AIS yields the trajectories of vessels, and this large dataset collectively generated by thousands of ships in the sea is critical for understanding vessel navigational behaviors. In the past few years, the AIS trajectory datasets are under focus for both research and practical applications. The major applications of AIS Data are: (a) analyzing maritime traffic patterns \cite{pallotta2013vessel,mazzarella2017novel} and collision avoidance \cite{harati2007automatic,statheros2008autonomous,mou2010study}, (b) vessel behavior study\cite{bomberger2006associative,rhodes2007probabilistic} and anomaly detection \cite{ristic2008statistical,mascaro2014anomaly}, (c) knowledge extraction\cite{felski2012information,harati2007automatic,mazzarella2014discovering,liu2014knowledge}and pattern mining \cite{zheng2015trajectory,vespe2016mapping}, and (d) trajectory clustering\cite{li2017dimensionality,liu2014knowledge} and predictions \cite{last2014comprehensive,skauen2016quantifying,pallotta2014context}.

Although the AIS data is useful for various purposes, its incompleteness or the presence of a large number of missing segments in its trajectories subdues its applicability. These missing segments are created either due to vessel inactivity, device failure, or inclement weather conditions at the point of data collection. In the past few years, this problem has led to the emergence of a new stream of studies for Trajectory Reconstruction, and several approaches are proposed to resolve this issue. Jie et al. \cite{jie2017novel} use bilateral filtering on the dynamic information of AIS for smoothening the trajectories and performing interpolations. \cite{zhao2018ship} focuses on improving the general applicability of AIS data by improving the quality of spatial and temporal data. \cite{sang2015novel} segments the trajectories based on their shape as line, curve, or arc and uses linear fitting for interpolation. In general, all these methods linear in nature and thus are only able to reconstruct easily inferable or relatively linear missing segments. Several researchers have also attempted to use Neural Network (NN), \cite{chen2019application} utilizes LSTM, and proposed a Neural Network-Based automatic reconstruction of Missing Vessel Trajectory Data. Although NN based methods generalize well for missing segments of short lengths, they fail to reconstruct segments of longer lengths. Also, these methods miss an opportunity to utilize patterns from trajectories of other objects present in the local vicinity of the target trajectory with the missing segments. Lastly, all the traditional methods, as well as NN-based methods, produce a single inference and do not account for its uncertainty. While the probabilistic inferences of missing observations incorporating uncertainty could be particularly useful within probabilistic pattern mining frameworks.

Mobility patterns are often repeated within the same or across different users trajectories. While mobility of objects (people, vehicles, ships) is often constrained of follows a number of (known/unknown) common rules. For ships, they often follow specific routes while navigating in the sea; thus, for any given trajectory, there may exist similar trajectories of the same or other vessels traversing the same route. These similar trajectories can be used to reconstruct missing gaps present in the given trajectory. 

We base our method on this principle and propose to impute the missing segment in the target trajectory by extracting and ensembling patterns from similar trajectories present in the spatial vicinity. We achieve this objective in three steps. (a) First, we select candidate trajectories in close proximity using box-similarity approach (b) Second, we calculate relevance weights for candidate patterns using similarities between the known portions of the target trajectory and segments corresponding to it within the candidate trajectories. (c) Third, we align all these candidate patterns and ensemble them using relevance weighting to generate the inference for the missing pattern.

We illustrate the proposed method applying it to AIS data and evaluate its performance to reconstruct missing segments of variable time and distance gaps in comparison with traditional functional interpolation methods, used as a baseline.

\section{AIS Dataset and Trajectory Processing}
The attributes of AIS data can be broadly categorised into three categories \cite{international_maritime_organization_2003}
: Static Data (vessel name and identity), Dynamic Data (position, heading, speed etc.), and Voyage Data (Draught, Cargo Description, Destination etc.). Last et al. \cite{last2014comprehensive}
 provides a comprehensive analysis and understanding of Automatic Identification System (AIS) data. To conduct this study we use the raw dynamic AIS Data \cite{vessel_traffic_data}
collected by the United States Coast Guard and provided by The Bureau of Ocean Energy Management (BOEM) and the National Oceanic and Atmospheric Administration (NOAA). The chunk of AIS data we used corresponds to U.S. and international waters of UTM Zone 18 \cite{vessel_traffic_data}.
This chunk of data contains AIS data and continuous positional logs of 7209 vessels for the month of July 2017. Overall, the data has 35 million positional points.

The raw AIS data is the continuous positional logs of vessels for the entire considered period; thus, it is noisy. It includes the cases of vessels being idle and yet transmitting their position as well as significant gaps where the positional points are missing. For this study, we clean these natural inconsistencies present in the AIS data and obtain clean segments of the trajectory that can be treated as consistent movements.

\begin{figure}[H]
  \centering
  \begin{tabular}{@{}ccc@{}}
    \includegraphics[width=.9\linewidth]{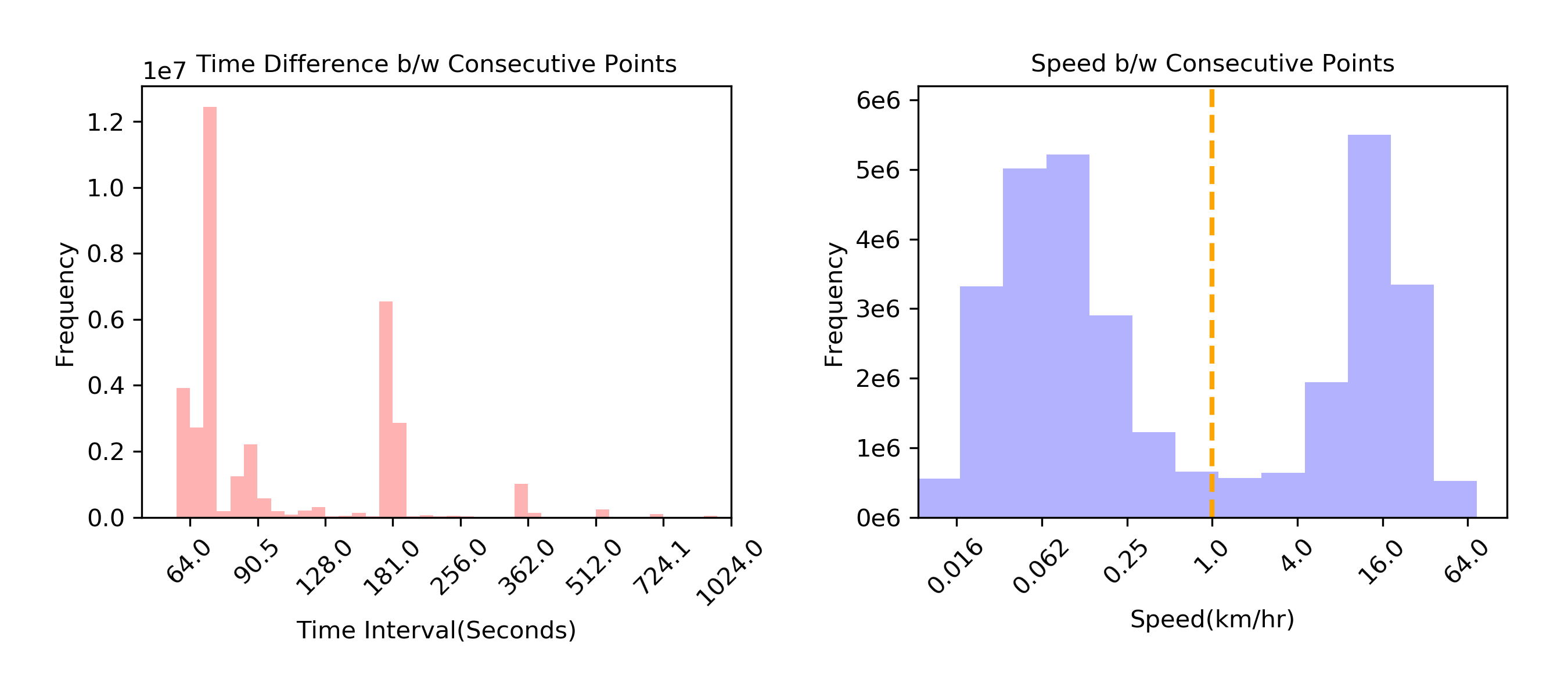} \\[\abovecaptionskip]
  \end{tabular}
  \caption{Distributions of Time Differences and Speeds between consecutive points of raw AIS trajectories.}
  \label{HistTimeDist}
\end{figure}
In order to eliminate points corresponding to the vessel's idle period, we use the distribution of vessel estimated speed between any two consecutive points. This distribution is shown in figure \ref{HistTimeDist}. The distribution is bimodal with a clear cut-off at 1km/hrs. Whereas the distribution of speed above 1km/hr follows the normal navigational behavior, speeds below it are unnaturally low and are likely to represent stationary periods. Thus, if in any section of the trajectory the speed between subsequent points is lesser than 1km/hr continuously for a long period, we mark the section as idle and split the trajectory at the beginning and end of this section. 

Next, we consider the distribution of time differences between the consecutive points.  This distribution peaks at 60,180, and 300 seconds. These time marks correspond to the frequency generally at which the AIS devices in the vessels transmit the data. 
Thus, if the time difference between any set of two consecutive points is higher than two or more times of the typical interval, we consider a missing segment between these points. Again to eliminate this missing gap, we split the trajectories at these points. While cleaning, we also found that 5\% of points in our dataset had an adjacent missing segment at one of their ends.

After forming the trajectories by eliminating noisy points, we performed cleaning at the trajectory level to eliminate noisy trajectories. Again, we use distributions of Time, Speed, and Distance at the trajectory level to execute this cleaning. Ultimately, we obtained 81K clean trajectories with 8.1 million positional points. The statistics corresponding to these clean trajectories could be found in table \ref{TimeSpeedDistribution1} and table \ref{TimeSpeedDistribution2}.

To develop an evaluation framework for testing the methods discussed in the next section, we create a sample set of 600 trajectories and induce synthetic missing segments in them. While creating this sample set, we controlled the time, distance and curvature of both the trajectories and the missing segments to have equal representation of linear/non-linear shapes and short/long lengths of the missing segments.

\begin{table}[H]
\centering
\caption{Distribution of attributes between consecutive points of clean trajectories}
\label{TimeSpeedDistribution1}
\begin{tabular}{@{}llll@{}}
\toprule
     & \multicolumn{3}{l}{Distribution of Attributes for Points (8183942 total)} \\ \midrule
     & Time(Seconds)         & Distance(Kms)         & Speed(Kms/hr)         \\ \cmidrule(l){2-4} 
mean & 75.30                 & 0.39                  & 18.91                 \\
std  & 19.75                 & 0.52                  & 40.11                 \\
25\% & 66.00                 & 0.26                  & 12.88                 \\
50\% & 70.00                 & 0.35                  & 16.59                 \\
75\% & 80.00                 & 0.47                  & 22.94                 \\ \bottomrule
\end{tabular}
\end{table}

\begin{table}[H]
\centering
\caption{Distribution of attributes at trajectory level for clean trajectories}
\label{TimeSpeedDistribution2}
\begin{tabular}{@{}lllll@{}}
\toprule
     & \multicolumn{3}{l}{Distribution of Attributes for Trajectories (81223 total)} &          \\ \midrule
     & Time(Seconds)        & Distance(Kms)        & Speed(Kms/hr)        & Number of Points \\ \cmidrule(l){2-5} 
mean & 7586.71              & 39.37                & 19.60                & 100.76   \\
std  & 14553.14             & 81.60                & 9.35                 & 202.61   \\
25\% & 1774.50              & 8.45                 & 13.06                & 23.00    \\
50\% & 2981.00              & 15.66                & 16.47                & 39.00    \\
75\% & 6199.00              & 32.25                & 22.89                & 78.00    \\ \bottomrule
\end{tabular}
\end{table}

\section{Alignment and Similarity between Trajectories}

\begin{figure}[H]
  \centering
  \begin{tabular}{@{}ccc@{}}
    \includegraphics[width=0.85\linewidth, height = 140pt]{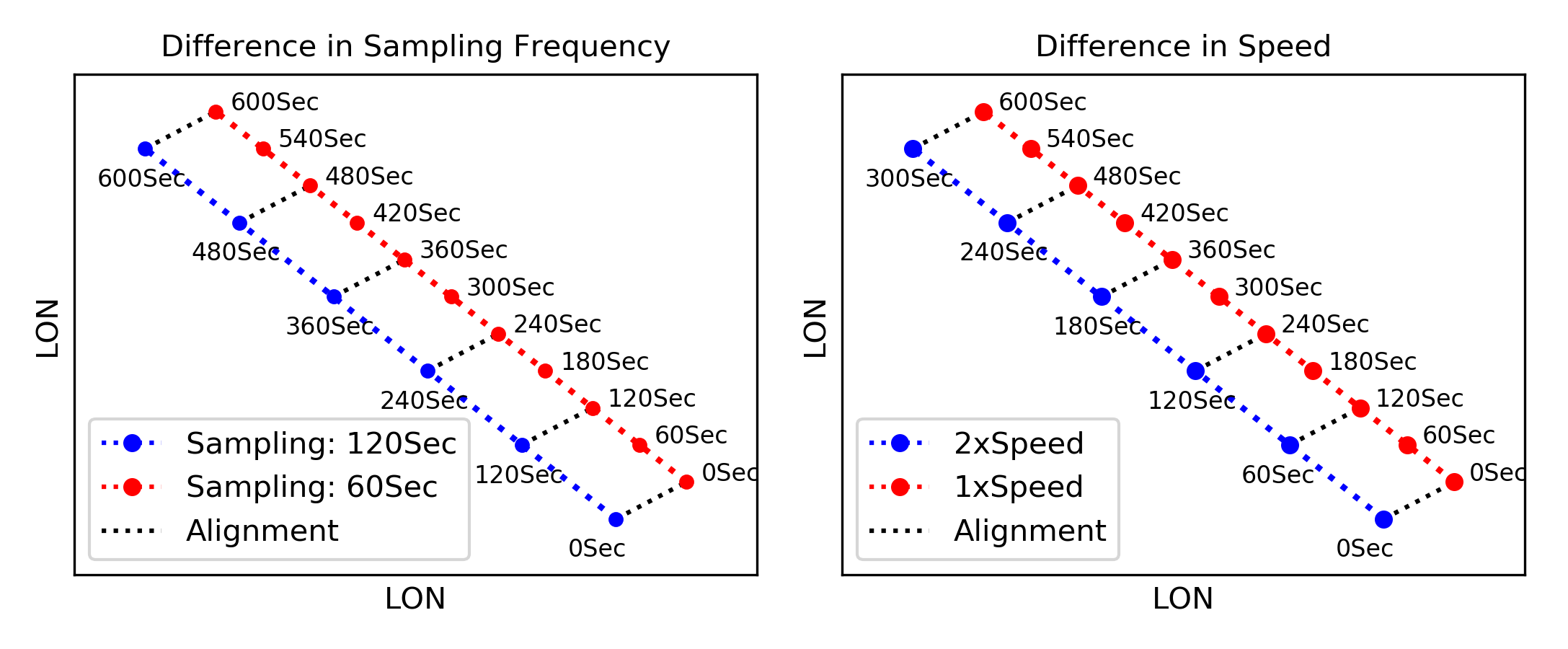} \\[\abovecaptionskip]
  \end{tabular}
  \caption{Illustration of similar trajectories with different number of points originating due to the difference in (a) sampling frequency and (b) speed of vessel }
  \label{comparison}
\end{figure}

Trajectory can be considered as a sequence of n points with temporal attribute timestamp (t) and positional attributes Latitude (LAT) and Longitude (LON).  When we compare two trajectories, there can be a difference between the number of points in them. This difference originates from two factors: (a) difference in time frequency at which the two trajectories were sampled or (b) difference in the speed of the vessels. Illustration of these factors is shown in figure \ref{comparison}.

To work with pairs of Trajectories, we need to find methods for (a) aligning the points and (b) calculating similarities between the sequences of location observations. There are multiple methods for calculating similarities between the two trajectories, the most used metrics are based on Euclidean distance\cite{jonker1980rounding},
LCSS (Longest common subsequence) \cite{kearney1990stream},
Hausdorff distance, Frechet distance, and DTW \cite{soong1988use}.

Distance measures like Euclidean and Manhattan assume prior alignment and calculate distances between paired sets of points matching the two trajectories. The mean or sum of these distances can be used as the final similarity measure. The requirement of prior alignment makes it difficult to use these methods with real-world trajectories. Another method, Longest Common Sub-Sequence (LCSS), considers only the distances between point pairs that have a distance below a certain threshold. The limitation of the threshold restricts the applicability of this method when the trajectories are diverging. Hausdorff and Frechet distances are based on calculating the maximum of minimum distances between the sets of points in the different trajectories. Frechet distance is a modification of Hausdorff and is more suitable for trajectory similarity as it also considers time sequence while aligning the sets of points i.e., any point in trajectory \textit{A} will not be paired up with any of the points in trajectory \textit{B} which were already paired up back in time.

Dynamic Time Warping (DTW) is a widely used method for finding similarity between temporal sequences of varying speeds. It matches every point in either of the trajectories with one or more points in the other and finds the most optimal match. DTW uses distance matrix and dynamic programming in a time-efficient way to return the most optimal match. Previously, Li et. al \cite{li_liu_liu_xiong_wu_kim_2017} 
have used DTW to find similarity between the trajectories, for developing their dimensionality reduction based multi-step clustering method. DTW fulfills both the match and similarity measure requirements of our study.

\section{Methodology}
Earlier in this paper, we defined trajectory as a sequence of points. Often inclement weather conditions or AIS device inactivity or failure leads to a missing set of points from the trajectories. 
Our study's objective is to infer a sequence of points that can most appropriately reconstruct the unobserved segment of the original trajectory.

A straightforward linear solution to this problem is to use the very last known points before and after the missing segment and interpolate a straight line between them. Hereafter, we will refer to this method as linear interpolation. A modification on top of the linear interpolation could be to use the last sets of points leading and trailing the missing segment and fit a polynomial function to use its predictions for reconstruction. In this study, we use both of these methods as our baseline for comparison. Similar to \cite{chen2019application}, we use cubic function as the baseline polynomial function, and hereafter we will refer to this as cubic interpolation. The number of known points to use for fitting cubic function is a hyper-parameter for this approach. Using a small number of points can lead to high variance, and using the entire set of known points may create high bias. Thus, we experimented with the different number of known points for fitting the cubic spline. The cubic interpolation performed best when using 15 known points on both ends of the missing segment, and we use this version for performance comparison with our proposed method.

In both of the above methods, we are using information present within the target trajectory having the missing segment. An alternative approach could be to also use the similar trajectories present in the spatial vicinity to provide likely patterns for reconstructing the observations gap. An ensemble of these patterns using a set of suitable relevance weights could lead to improved predictive performance. Our proposed method is based on this idea and performs in three stages: (a) Filter/select candidate trajectories, (b) Calculate relevance weights, (c) Align and ensemble patterns.

\begin{figure}[H]

  \centering
    \includegraphics[width=.9\linewidth]{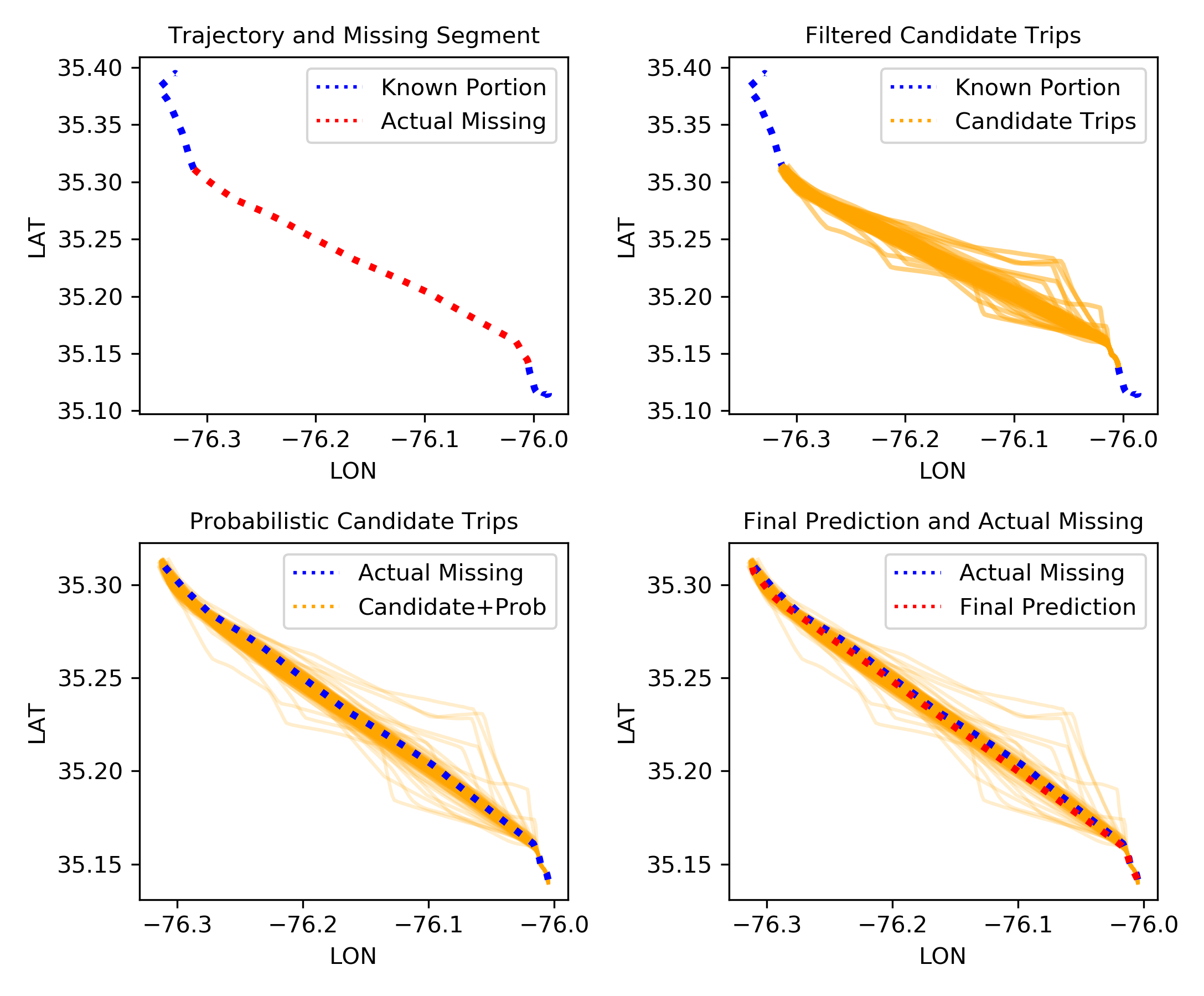}
\caption{The Three Stages of Pattern Ensembling}
  \label{methodology}
\end{figure}

One way to select similar trajectories is to compare the known portions of the target trajectory with every other available trajectory, calculate the similarities between them, and pick the top ones having the highest similarity score. The number of similarity score calculations required here will be equal to the number of reference trajectories and will grow with the size of the reference set, thus becoming computationally expensive. A much faster way could be to use spatial box constraints around the known points and filter only those trajectories which are traversing through them. To keep our method comparable with the cubic baseline, we use points belonging to 15-time steps before and after the missing segment and place a bounding box spatial constraints with sides of length 1Km on top of them. The selection of these points and the type of spatial constraints can be further improved for optimal performance.

After selecting the set of candidate trajectories, we calculate their relevance weights based on their relevance to the target trajectory and presumed ability to reconstruct the observations gap. To get these weights, we calculate the similarity between the known portions of the target trajectory and the segments corresponding to it in the candidate trajectories. Here, we use the DTW similarity score. To generate relevance weights for these candidate patterns with the mass summing up to 1, we use the softmax function from the multinomial classification on the set of similarity scores and redistribute the relevance weight amongst all the candidates.

In the last step, we align and ensemble the candidate patterns using their relevance weights to predict the final sequence for reconstructing the missing observations. Again, we use DTW for aligning points from different candidate segments. Note, DTW can only align two sets of points at a time, whereas we can have more than two candidate patterns for reconstruction. Thus, we start by picking two patterns with maximum relevance weights, align them using DTW, and average each pair of points based on the candidate's weights to create an ensembled pattern. This new ensembled pattern has a weight equivalent to the sum of the relevance weights of the candidate patterns it ensembled. Next, we pick the third candidate pattern and its weight, ensemble it with the already assembled pattern, and iteratively repeat this until all the candidate patterns are covered. One can assume that we can directly use the candidate pattern with the highest weight to reconstruct the missing gap. This method could work well in general cases, but due to its over-reliance on a single pattern, it can fail when the candidate patterns are not clearly relevant. In general, ensembling averages out inconsistencies in individual candidates and leads to increased inferential performance.

\section{Results and Discussion}

\begin{figure}[H]
  \centering
    \includegraphics[width=.9\linewidth,height=150pt]{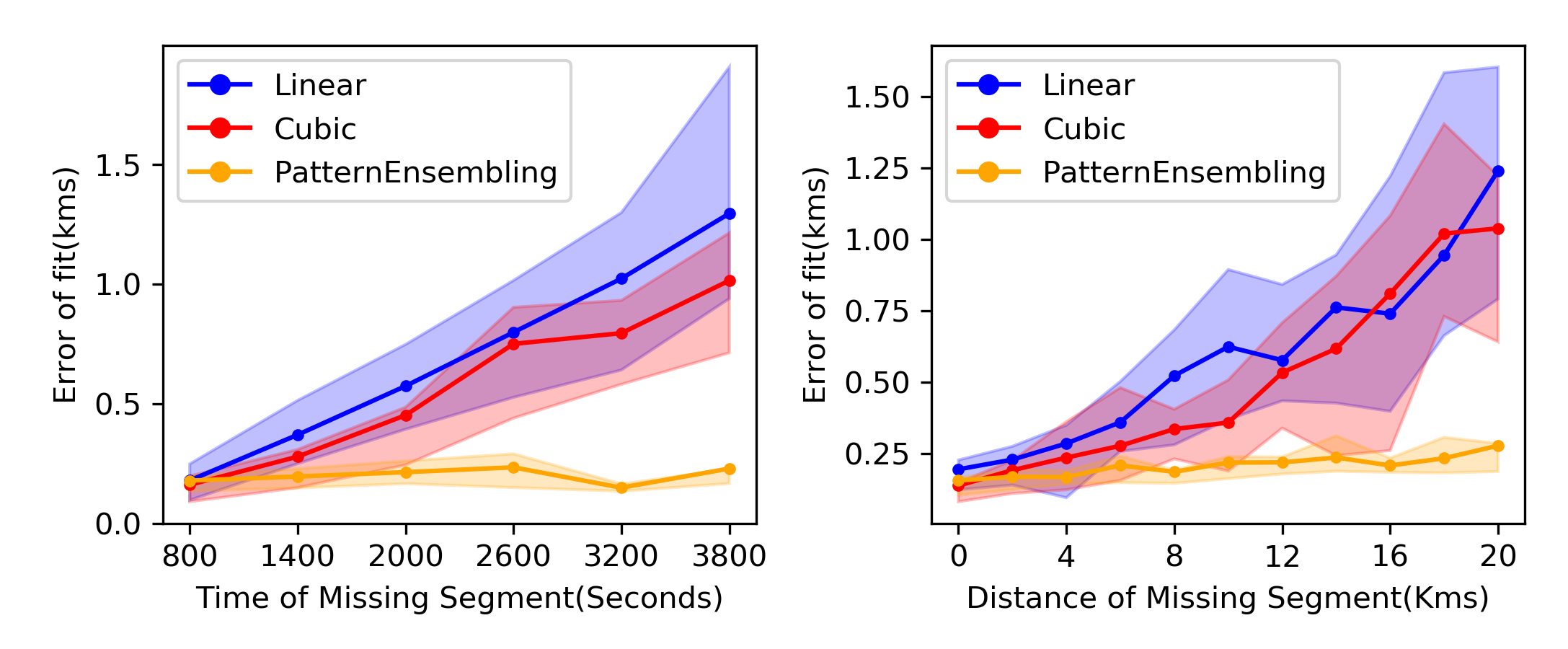}
\caption{Comparison of fit accuracy of Pattern Ensembling and Functional Interpolation with respect to time and distance of missing segments.}
    \label{linecharts}
\end{figure}

We evaluate the performance of the method for reconstructing the segments using the sample set created earlier with the induced missing segments. First, we align the prediction and the actual missing segment, 
next 
we calculate the geospatial distances between the paired sets of points in Kms and after that take their mean to get an average error of the missing segment inference. We conduct a quantitative study to evaluate the effect of increasing time/distance length of the missing segments on the different interpolation methods' performance. Moreover, we conduct a qualitative study to understand the role and impact of the the shape of trajectories or missing segments on the segment reconstruction performance.

The total of 111 samples out of 600 generated lack any suitable candidate trajectories in their spatial vicinity; thus, pattern ensembling does not generate any inference for those cases. The availability of a candidate trajectory is dependent on the tightness of spatial constraints used for filtering. The number of those cases could be reduced by optimizing the filtering strategy used in the first stage.

The mean error of the fit over the entire sample set for linear, cubic and the pattern ensembling interpolations is 0.478km, 0.403km, and 0.197km respectively. The significantly lower mean error for pattern ensemble interpolation highlights its superior performance over the other methods. To better understand this improved performance, we sort and bucket the samples based on the time and distance of their missing segment, and plot the mean errors with respect to those as shown on the Figure \ref{linecharts}. The plot clearly demonstrates that the fit error for linear and cubic interpolation increases with increased time and distance of the missing segment, whereas for the pattern ensembling this error remains consistently low. Functional interpolation methods assume that the trajectory follows certain analytic shape; while this assumption might not misrepresent reality when the length of the missing segment is relatively low, however when the length increases, for real trajectories, this assumption starts breaking down and large errors of fit accumulate.
On the contrary, the pattern ensembling method draws the patterns from other real-world trajectories and thus efficiently reconstructs the complex shapes of the real-world trajectories, undiscoverable for the functional interpolation. This feature enables the pattern ensembling method to keep its predictions close to the original shapes even for the longer missing segments.

\begin{figure}[H]
  \centering
  \begin{tabular}{@{}c@{}}
    \includegraphics[width=11cm]{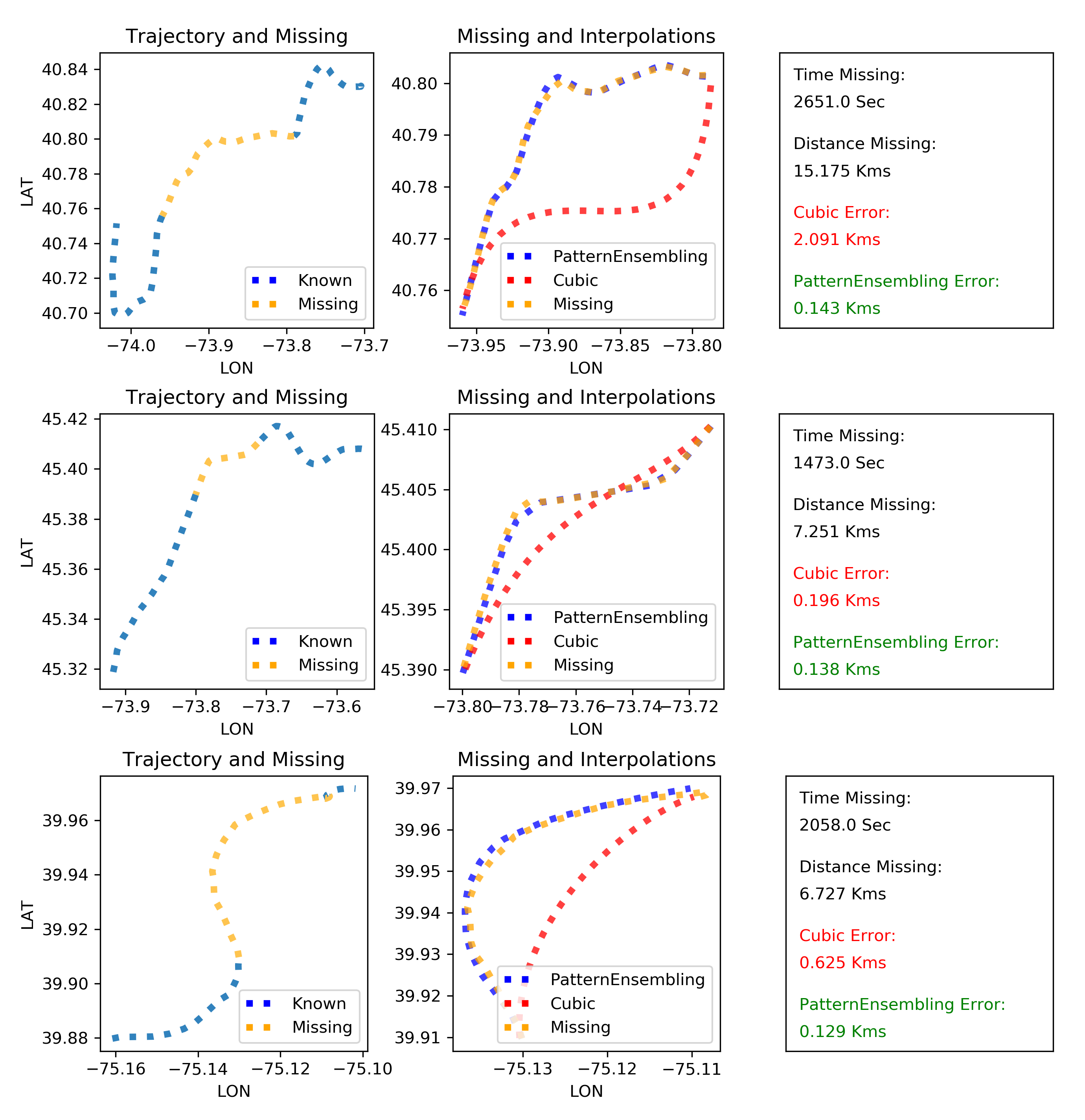} \\[\abovecaptionskip]
    \end{tabular}
    \caption{Cases with more complex shapes and geometries illustrating comparison of Cubic and Pattern Ensembling interpolations}
  \label{results}
\end{figure}

Our qualitative study of cases with complex shapes of trajectories reveals three more benefits of pattern ensembling over the functional interpolation methods. Figure \ref{results} illustrates these cases on specific examples.
First, pattern ensembling is robust against curvature and complexities of natural trajectories and provides reconstructions with minimal error, whereas the other methods fail to do so. Second, the baseline methods often fail to generalize, and the error accumulates to a high number when the missing segments are long. Third, the baseline interpolation methods overfit
the known data and are prone to incurring a massive off-shift in their prediction whereas, in our developed method, the probabilistic ensemble of different patterns helps to avoid overfitting and generalizes well.

\section{Conclusions}

In this paper, we consider the problem of missing segments in spatial trajectories data. As mobility patterns are repetitive in nature and often reproduced by the same or different users,
we proposed a new pattern ensembling method to reconstruct these missing segments. It effectively utilizes similar candidate patterns from local vicinity and probabilistically ensembles them to produce the reconstructed segment.

The three step approach to filter candidates, calculate their probabilistic relevance weights, and ensemble the inferences was successful in reconstructing missing segments in AIS trajectories. 
The empirical results demonstrate the superior performance of the proposed method for robustly reconstructing the missing segments compared to functional interpolation. The advantage of the pattern ensembling method is particularly noticeable while reconstructing segments of higher length/duration and more complex geometry. 

The approach can help recovering missing locations of mobile objects, such as people, vehicles or vessels and aid trajectory mining by providing evenly sampled trajectory representations.

\section{Acknowledgements.} The authors thank the US National Geospatial Intelligence Agency for supporting this work and further thank Christopher Farah for stimulating discussions and useful feedback.

\bibliographystyle{unsrt}
\bibliography{bibliography.bib}

\end{document}